\documentclass[reprint,superscriptaddress,aps] {revtex4-1}
\usepackage{graphicx}
\usepackage{hyperref}
\hypersetup{hidelinks}
\usepackage{amsmath}
\usepackage{color}

\begin{document}


\title{Self-synchronization of thermal phonons at equilibrium}

\author{Zhongwei Zhang}
\affiliation{Center for Phononics and Thermal Energy Science,\\
School of Physics Science and Engineering, Tongji University, 200092
Shanghai, PR China}
\affiliation{China-EU Joint Lab for Nanophononics, Tongji University, 200092 Shanghai, PR China}
\affiliation{Institute of Industrial Science, The University of Tokyo, Tokyo 153-8505, Japan}

\author{Yangyu Guo}
\affiliation{Institute of Industrial Science, The University of Tokyo, Tokyo 153-8505, Japan}

\author{Marc Bescond}
\affiliation{Laboratory for Integrated Micro and Mechatronic Systems, CNRS-IIS UMI 2820, The University of Tokyo, Tokyo 153-8505, Japan}

\author{Jie Chen}
\email{jie@tongji.edu.cn}
\affiliation{Center for Phononics and Thermal Energy Science,\\
School of Physics Science and Engineering, Tongji University, 200092
Shanghai, PR China}
\affiliation{China-EU Joint Lab for Nanophononics, Tongji University, 200092 Shanghai, PR China}

\author{Masahiro Nomura}
\email{nomura@iis.u-tokyo.ac.jp}
\affiliation{Institute of Industrial Science, The University of Tokyo, Tokyo 153-8505, Japan}

\author{Sebastian Volz}
\email{volz@iis.u-tokyo.ac.jp}
\affiliation{China-EU Joint Lab for Nanophononics, Tongji University, 200092 Shanghai, PR China}
\affiliation{Laboratory for Integrated Micro and Mechatronic Systems, CNRS-IIS UMI 2820, The University of Tokyo, Tokyo 153-8505, Japan}

\date{\today}

\begin{abstract}

Self-synchronization is a ubiquitous phenomenon in nature, in which oscillators are collectively locked in frequency and phase through mutual interactions. While self-synchronization requires the forced excitation of at least one of the oscillators, we demonstrate that this mechanism spontaneously appears due to the activation from thermal fluctuations. By performing molecular dynamic simulations, we demonstrate self-synchronization of thermal phonons in a platform supporting doped silicon resonators. We find that thermal phonons are spontaneously converging to the same frequency and phase. In addition, the dependencies to intrinsic frequency difference and coupling strength agree well with the Kuramoto model predictions. More interestingly, we find that a balance between energy dissipation resulting from phonon-phonon scattering and potential energy between oscillators is required to maintain synchronization. Finally, a wavelet transform approach corroborates the generation of coherent thermal phonons in the collective state of oscillators. Our study provides a new perspective on self-synchronization and on the relationship between fluctuations and coherence.

\end{abstract}

\pacs{Valid PACS appear here}
\maketitle


\section{\label{sec:level1}Introduction}

Synchronization of a population of coupled oscillators is a common phenomenon in nature, as observed in a wide range of physical and biological systems \cite{RN1774,RN1767,RN1691,RN1648,RN1752,RN1643,RN1700,RN1751}. Through mutual interactions, oscillators are self-organized into a collective motion, in which all synchronized units are locked to a single frequency and phase \cite{RN1774,RN1775,RN1648,RN1766}. Because of collective behaviors, synchronization in many fields has attracted intensive attention, for example to achieve coherent operation of micromechanical oscillators in optomechanics \cite{RN1637,RN1642,RN1648,RN1711,RN1643}. Usually, synchronization is understood as a stationary state sustained by external forces \cite{RN1774,RN1724,RN1751}, in which the rate of change of the entropy $S$ can be properly decomposed as \cite{RN1724,RN1723,RN1697}

\begin{eqnarray}
\centering
\frac{dS}{dt}=\Pi -\Phi ,
\label{eq_0}
\end{eqnarray}

\noindent where $\Pi$ is the entropy production due to irreversible processes inside the system and $\Phi$ is the entropy flux between the system and its environment. In the nonequilibrium stationary state of synchronization, i.e. $\Pi=\Phi$, $\Pi$ corresponds to thermal dissipation, and $\Phi$ results from the driving action which stabilizes the stationary state \cite{RN1724,RN1700,RN1751}. When this driving action excites each oscillator independently, the resulting synchronization is named self-synchronization. In the following, ‘synchronization’ might be used as a shortened term for self-synchronization. As demonstrated by Zhang $et$ $al.$ \cite{RN1751} the synchronization of coupled molecular oscillators requires external driving action as an energy cost.

As principal energy carriers in solids, thermally activated phonons \cite{RN1764,RN870,RN1771,RN1230,RN1811} display a coherence that is analogous to the one of a collection of interacting oscillators in other physical and biological systems \cite{,RN1767,RN1691,RN1648,RN1752,RN1700,RN1751}. Consequently, the logical connection between synchronization and coherence has led us to postulate the possibility of self-synchronization activated by thermal phonons \cite{RN975,RN1768,RN1230,zhang2020generalized,RN1729,RN606,RN184}. 

By using direct molecular dynamic (MD) simulations, we will demonstrate passive self-synchronization in a resonator system due to its activation by thermal phonons produced by equilibrium fluctuations. In contrast to usual self-synchronized systems, we highlight the absence of external driving action here, that we express by the adjective ‘passive’. Synchronizations of frequency and phase are proven and investigated. The relevance of previous theoretical models is confirmed in the present frame of thermal phonon activated resonators. The effects of frequency difference, coupling strength, and temperature on synchronization are discussed. In addition, the generation of coherent thermal phonons after synchronization is also investigated through a wavelet transform approach. Proving self-synchronization establishes another framework for the understanding of the dynamics of thermal phonons and provides a new route for the generation of coherent thermal phonons.

\begin{figure*}[t]
\includegraphics[width=0.8\linewidth]{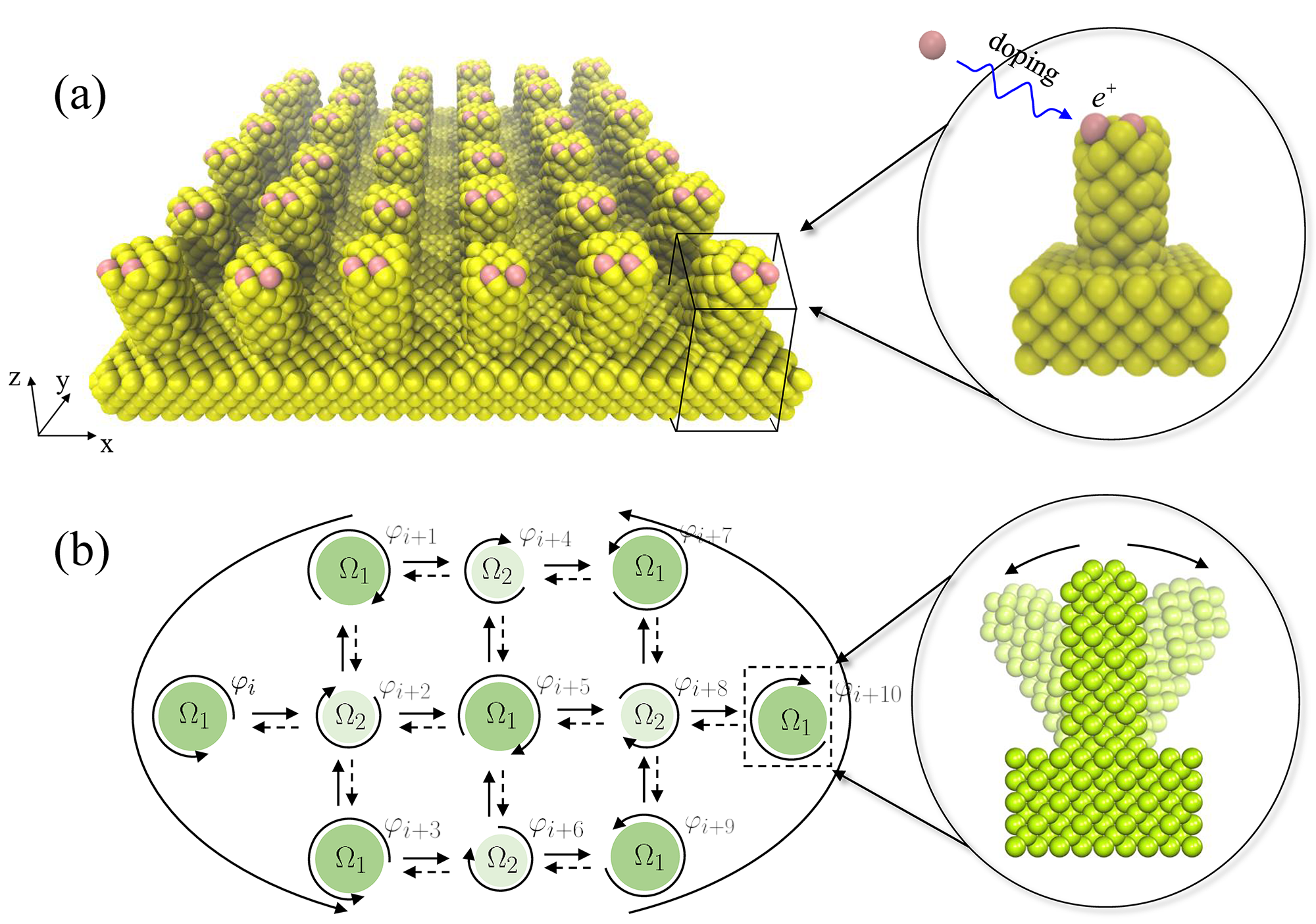}
\caption{Schematic figure of the doped silicon resonator system. (a) Pillared silicon membrane with electron doping (pink atoms) on the top of pillars. Pillars behave as resonators activated by thermal phonons. The zoom-in figure shows a unit of resonator. (b) Schematic figure of the frequencies ($\Omega$) and phase ($\varphi$) dynamics of resonators under mean-field coupling. The system consists of two type resonators with frequencies $\Omega_1$ and $\Omega_2$, and random phases $\varphi _{i,i+1,\cdots }$. The zoom-in figure shows the vibration of a pillar as a unit of the resonator.
}
\label{fig1}
\end{figure*}

\section{\label{sec:level1}Methodology}

To directly investigate the dynamics of thermal phonons, we adopt a silicon pillared membrane as the platform. Previous studies \cite{RN1649,RN1704} found that pillars on the surface of a membrane act as local resonators, and their predominant and resonance frequency is expectedly decreasing with the pillar size \cite{RN7,RN1647}. As shown in Fig. \,\ref{fig1}(a), the system is composed of a set of pillar resonators and a membrane. In a unit, the dimension of the membrane element is fixed to 2.18 nm$\times$2.18 nm$\times$1.09 nm, while the dimension of the pillars is 1.09 nm$\times$1.09 nm in x-y directions and the height varies from 2.4 nm to 3.3 nm. Thus, the distance between pillars is set to 2.18 nm. To achieve frequency difference between the resonators, we consider two intercalated types of pillars with different height, with corresponding frequencies $\Omega_1$ and $\Omega_2$. By extending the unit cells to a 16$\times$16 supercell, we obtain a resonator system containing two intrinsic frequencies and a distribution of randomly distributed phases $\varphi$ (see Fig. \,\ref{fig1}(b)). Two kinds of frequency differences are studied, $\delta \Omega=\left | \Omega_1-\Omega_2 \right |$, which are characterized by two different frequency differences, $\delta \Omega_I=$0.02 THz and $\delta \Omega_{II}$=0.04 THz. In addition, to introduce a mean-field coupling between resonators, a long-range electrostatic force is added by doping to the top of the pillars, as shown in Fig. \,\ref{fig1}(a). By varying the doping ratio, four systems are investigated, i.e. $\delta \Omega_{I}$ with doping ratio 0.2 $\%$ (I1) and 0.4 $\%$ (I2), and $\delta \Omega_{II}$ with doping ratio 0.2 $\%$ (II1) and 0.4 $\%$ (II2). Li $et$ $al.$ \cite{RN1753} found the dopants in a confined nanostructure can be treated as localized charges, which were simulated here.

We use classical MD simulations to study the dynamics of thermal phonons. The covalent Si-Si interaction is modeled by the Tersoff potential \cite{RN1007}. The electrostatic interaction between dopants is modeled by the standard Coulombic formula with a cutoff distance of 25.0 $\AA$, and longer-range interactions that exceed the cutoff are simulated by the $pppm$ kspace method \cite{RN1772}. Periodic boundary conditions are applied in the x and y directions. Compared to the strong covalent bonding, the long-range electrostatic interaction is several orders of magnitude smaller, which means that the intrinsic frequencies of resonators are not altered due to the electrostatic forces. All MD simulations are performed by using the LAMMPS package \cite{RN374} with a timestep of 0.35 fs. Firstly, the system is relaxed in the isothermal-isobaric (NPT) ensemble with $10^{5}$ steps. Then, the simulation runs over $2\times 10^{5}$ steps with constant temperature in the canonical (NVT) ensemble. During these two processes, the dynamics of thermal phonons is controlled by an external thermostat, here the Nosé–Hoover thermostat \cite{RN1773} is used. Then, the microcanonical (NVE) ensemble is carried out. During this process the evolution of thermal properties is analyzed.

\section{\label{sec:level1}Self-synchronization of thermal phonons}

\subsection{\label{sec:level1}Kuramoto model}

Previously, synchronization dynamics of oscillators in optomechanics and other classical systems have been well understood by using theoretical models, in particular the Kuramoto model \cite{RN1775,RN1774,RN1637,RN1751,RN1642}. The Kuramoto model provides a fundamental description of self-synchronization of thermal phonons. It describes a non-linearly coupled system of $N$ oscillator phases $\varphi _i$, with intrinsic frequency $\Omega_i$. For a two resonators system, the dynamics of the phase difference $\delta \varphi$ is described according to \cite{RN1775}

\begin{eqnarray}
\centering
\delta \dot{\varphi} =\delta \Omega  -2 K sin\left (\delta  \varphi \right ),
\label{eq_1}
\end{eqnarray}

\noindent where $\delta \Omega\equiv \Delta \Omega  $ is the difference between two eigenfrequencies, and $K$ is the reduced coupling constant between resonators. At frequency synchronization, one would find frequency difference $\delta \dot{\varphi}=0$, as $\dot{\varphi}$ here corresponds to a frequency. Thus, when the coupling constant $K$ exceeds the threshold $K  _{c}=\delta \Omega /2$ with $sin\left (\delta  \varphi \right )=1$, frequency synchronization is appearing. On the other side, the phase synchronization happens when the phase lag $\delta \varphi=0$, indicating that the coupling constant $K$ should be larger than the threshold, $K>K _{c}$. Obviously, the phase synchronization appears to be even harder to achieve than the frequency synchronization. In more realistic cases, however, the dynamics of frequency and phase become more complex. Heinrich $et$ $al.$ \cite{RN1637,RN1719} proposed to include an amplitude assisted dynamics to the phase evolution in the following form

\begin{eqnarray}
\centering
\delta \dot{\varphi} =\delta \Omega -\left [ Ccos\left ( \delta \varphi \right )+Ksin\left ( 2\delta \varphi\right ) \right ].
\label{eq1}
\end{eqnarray}

\noindent The constants are expressed by $C=\frac{k}{2M}\left | \frac{1}{ \Omega_{i}} -\frac{1}{ \Omega_{j}}  \right |$ and $K=\frac{k^{2}}{8M^{2}\gamma }\left ( \frac{1}{ \Omega_{i}^{2}} +\frac{1}{ \Omega_{j}^{2}}\right )$. Here, $k$ is the harmonic coupling between oscillators emanating from electrostatic force, $\gamma$ is the damping constant that results from anharmonicity, and $M$ is the mass of each oscillator. In the right-hand-side of Eq. (\ref{eq1}), the last two terms lead the dynamics of the phase and the frequency. The coupling $k$ contributes positively to drive the synchronization, but the anharmonicity $\gamma$ acts negatively as dissipation, which agrees well with the definition of Eq. (\ref{eq_0}). The Kuramoto model is widely verified in mechanical \cite{RN1774,RN1637,RN1642} and biological \cite{RN1700,RN1751} systems, its applicability to phonons remains to be discussed.

\subsection{\label{sec:level1}Self-synchronization in frequency}

\begin{figure}[t]
\includegraphics[width=1.0\linewidth]{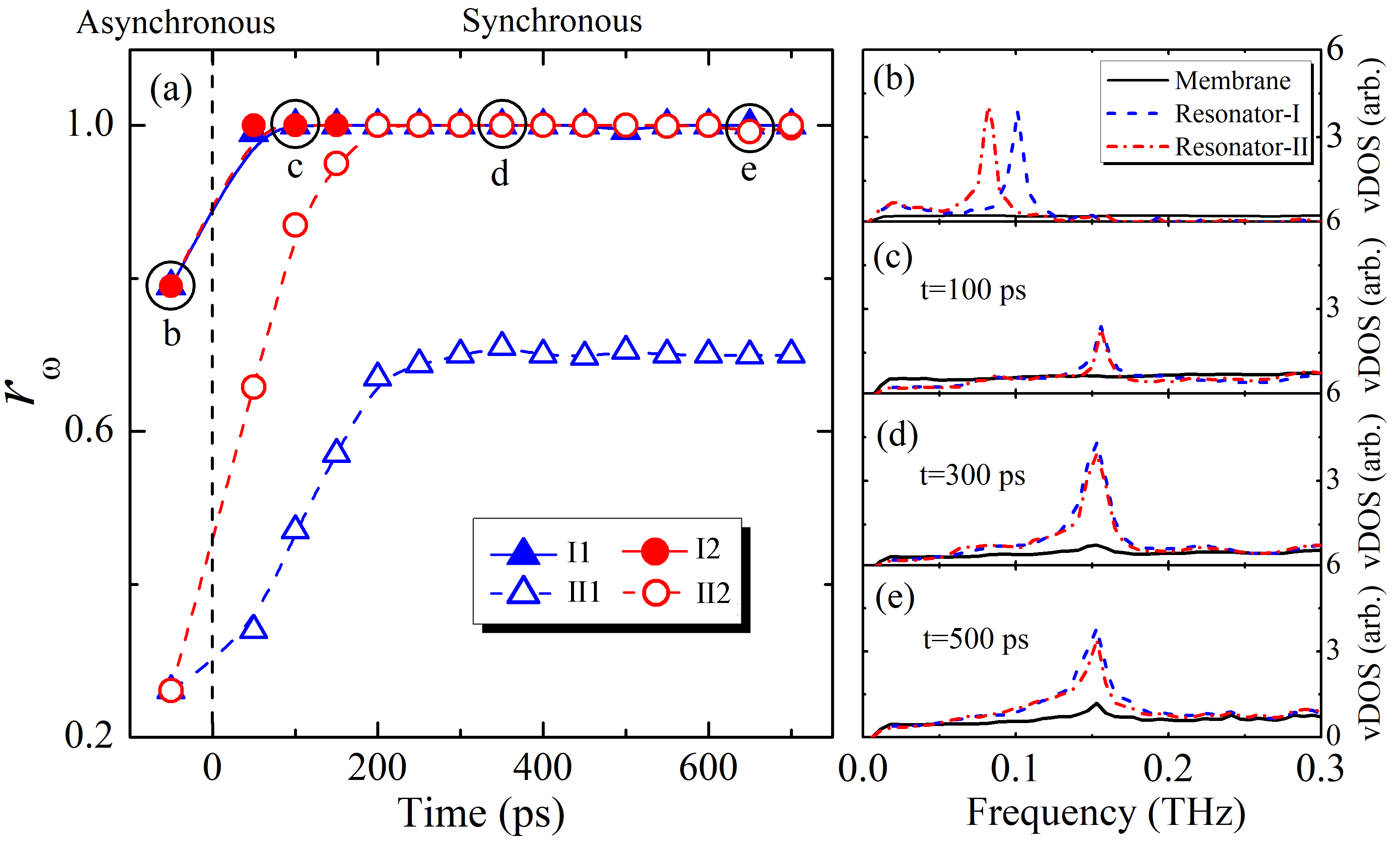}
\caption{Self-synchronization of thermal phonons in frequency. (a) Synchronization degree in frequency ($r_{\omega }$) versus evolution time for four systems, i.e. I1, I2, II1 and II2, at 100 K. (b-e) Vibrational density of states ($vDOS$) of membrane, resonator-I and -II for the I1 system. Times are also referenced as black circles in Fig. \,\ref{fig2}(a).
}
\label{fig2}
\end{figure}

The frequency information of the resonators is obtained from the vibrational density of states ($vDOS$),

\begin{eqnarray}
\centering
vDOS\left ( \omega  \right )=\frac{1}{ n_{a }}\sum_{i,\alpha }\left | \int_{t_{1}}^{t_{2}}\upsilon _{i,\alpha}\left ( t \right )e^{i\omega t} dt\right |^{2}.
\label{eq_2}
\end{eqnarray}

\noindent $\upsilon _{i,\alpha}\left ( t \right )$ refers to the atomic velocity of the $i$-th atom along $\alpha$ direction at time $t$, and $n_{a }$ is the number of atoms in the summation. The integration corresponds to the time interval from $t_{1}$ to $t_{2}$ over which the frequency information is averaged at the time ($t_{2}$-$t_{1}$)/2, and the $vDOS$ is averaged over atoms of resonators and membrane. For the undoped system, two different resonance frequencies are observed, as shown by the peaks of the $vDOS$ in Fig. \,\ref{fig2}(b), indicating the asynchronous state in the uncoupled system. The synchronization degree in frequency can be defined as 

\begin{eqnarray}
\centering
r_{\omega }=1-\frac{\Delta \Omega }{\bar{\Omega }},
\label{eq_3}
\end{eqnarray}

\noindent where, $\Delta \Omega$ is the frequency difference given by the peak of the time-dependent $vDOS$, and $\bar{\Omega }$ denotes the averaged frequency. $r_{\omega }=1$ means that the system is fully synchronized in frequency, while in the asynchronous state $r_{\omega }$ remains always smaller than 1.0 (see Fig. \,\ref{fig2}(a)). $r_{\omega }$ also depends on the intrinsic frequency difference.

As switching from NVT to NVE ensembles, the dynamics of the coupled resonators becomes unconstrained. The self-synchronization in frequency of thermal phonons rapidly emerges. As shown in Fig. \,\ref{fig2}(a), synchronization degree $r_{\omega }$ is increasing with the evolution time. For the I1 system, the thermal phonons are quickly synchronized to the same frequency, i.e. the synchronization frequency ($\omega_{s}$), which is manifested by the degenerated peaks in the $vDOS$ spectrum (see Figs. \,\ref{fig2}(c-e)). A comparison of phonon dispersion (0 K) between undoped and I1 systems shows that the weakness of electrostatic interactions between resonators has a negligible effect on the intrinsic vibration properties (See Fig. S1 in \cite{SM}). We can conclude that the observed frequency change and the degeneration of vibrational properties for different resonators are originating from the effect of self-synchronization by the thermal fluctuation. The synchronization of pillars also introduces the vibration of the membrane at the synchronization frequency $\omega_{s}$, as revealed by the increased amplitude of the membrane $vDOS$ in Fig. \,\ref{fig2}(e).

Moreover, the effect of intrinsic frequency difference and coupling strength between resonators on the frequency self-synchronization is revealing consistency with theoretical predictions. The Kuramoto model in Eq. (\ref{eq_1}) indicates that the self-synchronization in frequency can be enhanced by increasing the coupling strength or decreasing frequency difference. Correspondingly, the simulation results in Figs. \,\ref{fig2}(a) show that the system with small frequency difference ($\delta \Omega _{I}$) or with high doping ratio (0.4 $\%$) are more easily synchronized.

\subsection{\label{sec:level1}Self-synchronization in phase}

\begin{figure}[b]
\includegraphics[width=1.0\linewidth]{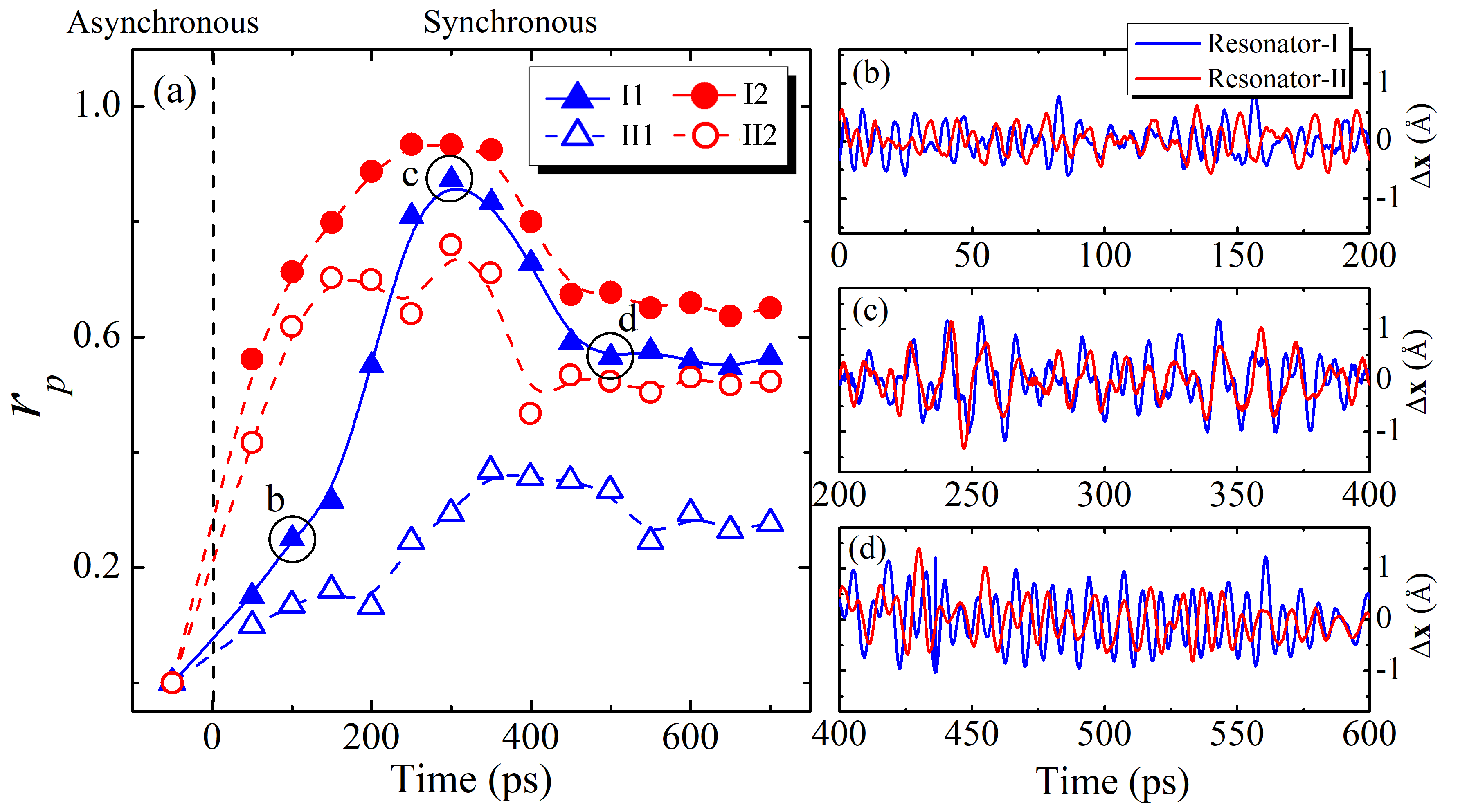}
\caption{Self-synchronization of thermal phonons in phase. (a) Synchronization degree in phase ($r_{p}$) versus evolution time for four systems, i.e. I1, I2, II1 and II2, at 100 K. (b-d) Averaged displacement of two neighbored resonators versus evolution time for the I1 system. Times are also referred as black circles in Figs. \,\ref{fig3}(a).
}
\label{fig3}
\end{figure}

On the other hand, resonators can also be synchronized in phase. The synchronization degree of phase ($r_{p}$) reads as \cite{RN1774}

\begin{eqnarray}
\centering
r_{p}e^{i\Theta}=\frac{1}{N}\sum_{i}e^{i\varphi_{i}},
\label{eq_4}
\end{eqnarray}

\noindent where $\Theta$ denotes the phase average. The $e^{i\varphi_{i}}$ term is calculated as the normalized displacement of the $i$-th resonator in the MD simulations, and the displacement is averaged over the atoms of each pillar. Before synchronization, the resonators in the undoped system exhibit uncorrelated dynamics (see Fig. \,\ref{fig3}(b)), as $r_{p}\approx 0.0$.

Under the free condition of the NVE ensemble, the transient process of phase synchronization is investigated in Fig. \,\ref{fig3}. Compared to the rapid and monotonous frequency synchronization, the phase synchronization displays a distinct behavior. As time evolves from 0 to 300 ps, the resonators are gradually synchronized and synchronization degree reaches the highest point around 300 ps. At the time of 300 ps, the resonators become almost fully synchronized, with for instance $r_{p}\approx 0.9$, in the case of the I2 system. The displacement dynamics of two resonators agree well with each other around 300 ps (see Fig. \,\ref{fig3}(c)), exhibiting an excellent collective synchronized state. However, after 300 ps, $r_{p}$ is decreasing with time, manifesting a de-synchronization process. In the last stage, $r_{p}$ remains at stationary state after 500 ps, indicating an only partial phase synchronization (Fig. \,\ref{fig3}(d)). 

	We find that the phase synchronization of thermal phonons also depends on the frequency difference between resonators and on the coupling strength due to doping in a similar way than for frequency synchronization. If we reduce the frequency difference and improve coupling strength, the phase synchronization can be enhanced. Moreover, compared to the synchronization of frequency, the phase synchronization is more difficult to achieve, especially the fully synchronized state, which agrees well with the prediction of Kuramoto model \cite{RN1774,RN1637}. Obviously, the phase synchronization process requires a deeper insight more specifically regarding its dynamics. Note that the coherent state or collective state of thermal phonons is always understood as locked in phase \cite{RN545,RN1438}. Thus, in the following, we mainly focus on the phase synchronization of thermal phonons.

\section{\label{sec:level1}Energy conversion driven self-synchronization}

\subsection{\label{sec:level1}Entropy change}

The self-synchronization leads thermal phonons from a disordered state to an ordered state. Apparently, this transition to a coherent motion indicates a reduction of entropy due to the reduction of state number, which raises the issue of the validity of the second law of thermodynamics. To provide an insight in the detailed mechanisms occurring during the self-synchronization of thermal phonons, the entropy is calculated from the MD simulations \cite{RN1695,RN1779} as

\begin{eqnarray}
\centering
S=\frac{k_{B}}{2}\mathrm{ln}det\left (\frac{k_{B}T e^{2}}{\hbar^{2}}\mathbf{M}\mathbf{\sigma } +\mathbf{1}\right ),
\label{eq_5}
\end{eqnarray}

\noindent where $k_{B}$ refers to the Boltzmann constant, $\hbar$ is the reduced Planck constant, $e$ is the Euler's number and $T$ corresponds to the temperature. $\mathbf{M}$ and $\mathbf{1}$ are the mass matrix and the unity matrix, respectively. $\mathbf{\sigma }$ is the covariance matrix of the coordinate fluctuations, with $\sigma _{ij}=\left \langle \left ( x_{i}-\left \langle x_{i} \right \rangle \right ) \left ( x_{j}-\left \langle x_{j} \right \rangle \right )\right \rangle$. The calculated entropy for the whole system and the resonator part for the I1 system are shown in Fig. \,\ref{fig4}(a). Before 300 ps, the entropy is continuously increasing, which corresponds to the synchronization process of Fig. \,\ref{fig3}(a), and resonators contribution is predominated by the entropy change due to the synchronization. Therefore, the self-synchronization of thermal phonons indeed follows the second law of thermodynamics. After 300 ps, the entropy of the resonators and the whole system reaches a stationary-like state with small fluctuations, but still in agreement with the second law of thermodynamics.

\subsection{\label{sec:level1}Energy conversion}

\begin{figure}[t]
\includegraphics[width=1.0\linewidth]{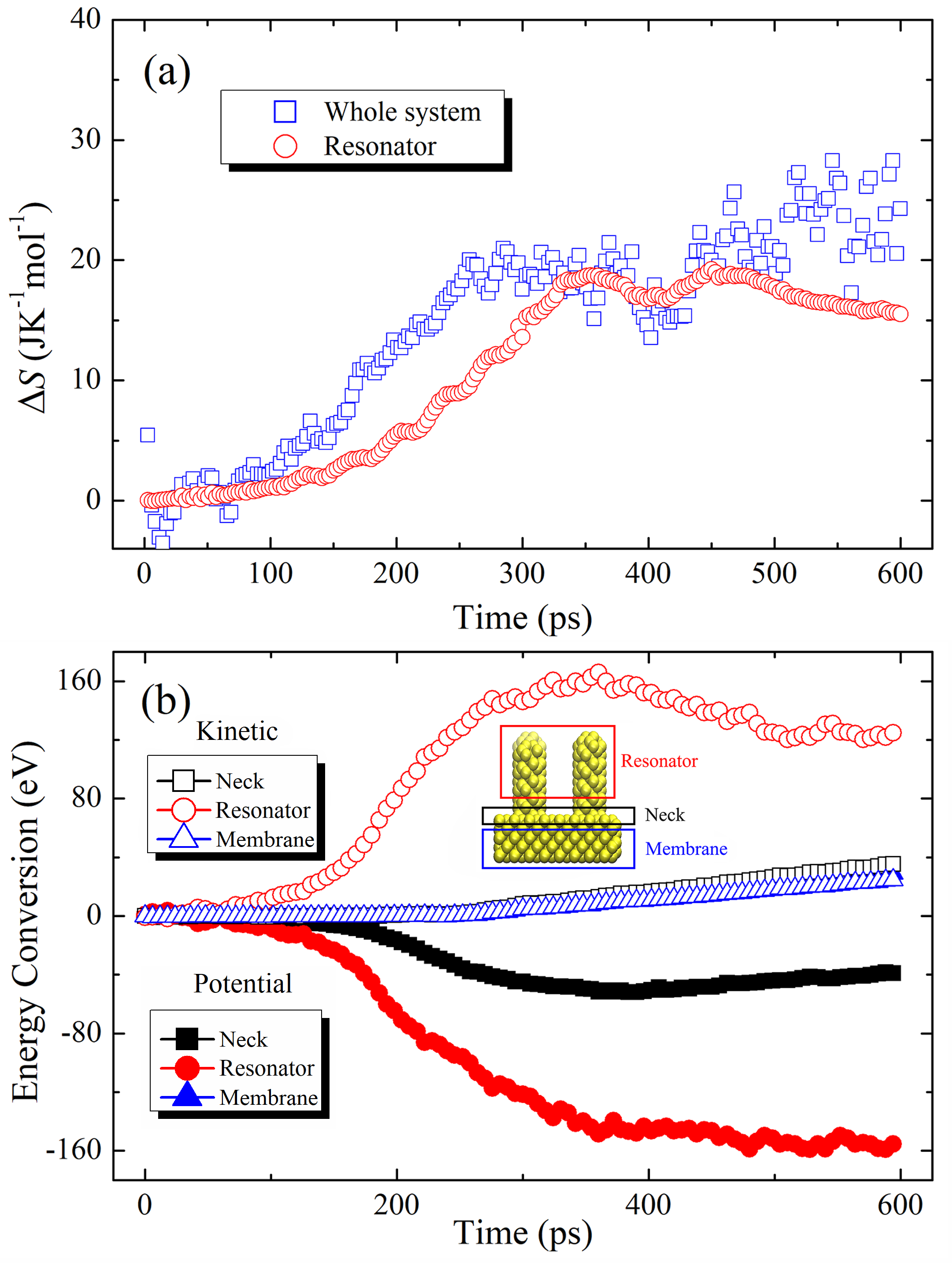}
\caption{Entropy change and energy conversion during self-synchronization of thermal phonons. (a) Entropy change of the whole system and of the resonators versus evolution time. (b) Kinetic energy and potential energy of neck, resonator and membrane regions versus evolution time of self-synchronization. The inset figure shows the definition of the neck, resonator and membrane regions. The I1 system is studied and its temperature is set to 100 K in both (a) and (b). 
}
\label{fig4}
\end{figure}

To further investigate the transformation during self-synchronization, we analyze the energy conversion in different regions of the system. Considering the possible strain in the neck region between pillars and membrane, we divide the system into three regions, i.e. neck, resonator and membrane, as shown in the inset figure of Fig. \,\ref{fig4}(b). The potential energy and kinetic energy is calculated respectively by summing atom energies, and those energies are excited from the thermal fluctuation. Fig. \,\ref{fig4}(b) shows that there is an obvious energy conversion between potential energy and kinetic energy, especially in the pillars. Before 300 ps, because of the principle of potential energy minimization \cite{RN1780}, the relative displacement between pillars is spontaneously reduced to minimize potential energy, leading the system from an asynchronous state to a synchronous state. The lowered down potential energy inside the resonators and neck region is converted into kinetic energy in the resonators. Therefore, in the coupled system, the minimization of potential energy drives the synchronization of thermal phonons, and also actually results in the increase of entropy. Expectedly, the total number of microscopic configurations is increasing with kinetic energy. In the studied systems, the potential energy gain is insufficient to fully achieve phase synchronization. Simultaneously, the energy communication between pillars and membrane, i.e. phonon-phonon scattering inducing energy dissipation and energy transfer, slightly increases energy in the membrane (see Fig. \,\ref{fig4}(b)), especially at synchronization frequency $ \omega _{s}$. The membrane motion is a necessary mechanism, absence of synchronization was indeed observed when the in-plane degrees of freedom were removed in the membrane.

\begin{figure}[t]
\includegraphics[width=1.0\linewidth]{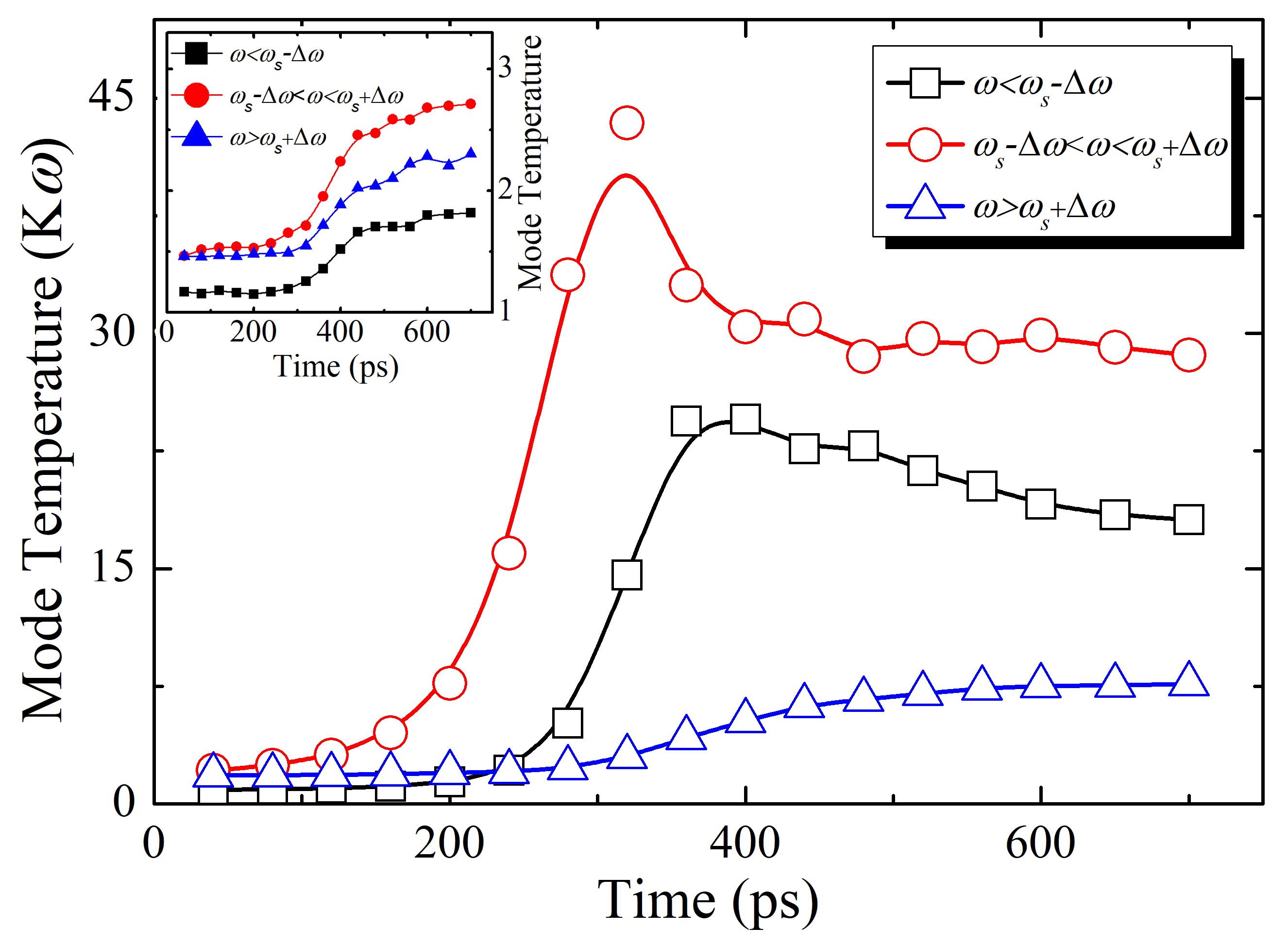}
\caption{Mode temperature evolution during self-synchronization of thermal phonons. (a) Mode temperature of frequency regions I) $\omega <\omega _{s}-\Delta \omega$, II) $\omega _{s}-\Delta \omega<\omega <\omega _{s}+\Delta \omega$ and III) $\omega _{s}+\Delta \omega<\omega$ in the resonators versus evolution time of self-synchronization for the I1 system at 100 K. The inset figure shows the mode temperature evolution in the membrane.
}
\label{fig5}
\end{figure}

However, the energy conversion cannot explain the de-synchronization and the stationary state after 300 ps. Considering the resonant nature of pillars, the synchronization of thermal phonons should be mode dependent, as shown in Fig. \,\ref{fig2}. We define the mode temperature \cite{RN1228} of the resonators for different frequency intervals,

\begin{eqnarray}
\centering
\tilde{T}\left ( t \right )=\frac{1}{k_{B}m n_{a } n_{\omega }}\sum_{i} \left | \int_{\omega_{1} }^{\omega_{2} }\upsilon _{i,\alpha}\left ( t \right )e^{i\omega t}  d\omega\right |^{2},
\label{eq_6}
\end{eqnarray}

\noindent where, $m$ corresponds to the mass of a silicon atom and $ n_{\omega }$ is the number of terms in the discrete summation. To study the mode dependent information, three integration intervals from $\omega_{1}$ to $\omega_{2}$ are calculated : I) $\omega_{1} =0;$ $\omega_{2} =\omega _{s}-\Delta \omega$, II) $\omega_{1} =\omega _{s}-\Delta \omega;$ $\omega_{2} =\omega _{s}+\Delta \omega$, and III) $\omega_{1} =\omega _{s}+\Delta \omega;$ $\omega_{2} =\infty$. $\Delta \omega$ is the frequency broadening of the peak at $\omega _{s}$. The calculated mode temperature is shown in Fig. \,\ref{fig5}. During self-synchronization before 300 ps, the minimized potential energy is mainly converted to the thermal energy around frequency $\omega _{s}$. This indicates that the synchronization in this coupled resonator system is emerging from the thermal phonons corresponding to the pillar resonance. In addition, the amplitude of thermal energy at frequency $\omega _{s}$ should be proportional to the synchronization degree.

Furthermore, Fig. \,\ref{fig5} shows that in the vicinity of 300 ps the thermal energy at frequency $\omega _{s}$ transfers to other modes inside the pillars presumably through phonon-phonon scattering, especially for the low frequency modes by an annihilation process, such as $\omega _{s} \rightarrow {\omega}'+{\omega}''$ \cite{RN932,RN82}. Accordingly, the energy at synchronization frequency is decreasing, indicating that the phonon-phonon scattering resistively contributes to the synchronization, which manifests the role of the entropy production $\Pi$ in Eq. (\ref{eq_0}). This dissipation is analogous to the energy cost of synchronization in biological systems \cite{RN1751}. We can observe the de-synchronization in phase from 300 ps to 500 ps (Fig. \,\ref{fig3}(a)) due to the decrease in synchronization energy (Fig. \,\ref{fig5}). It should be noted that the resonant vibrations in pillars are highly localized, i.e. with zero group velocity. Thus, we can find in the inset figure of Fig. \,\ref{fig5} that the increase of mode temperature in the membrane occurs with low amplitude.

\begin{figure}[b]
\includegraphics[width=1.0\linewidth]{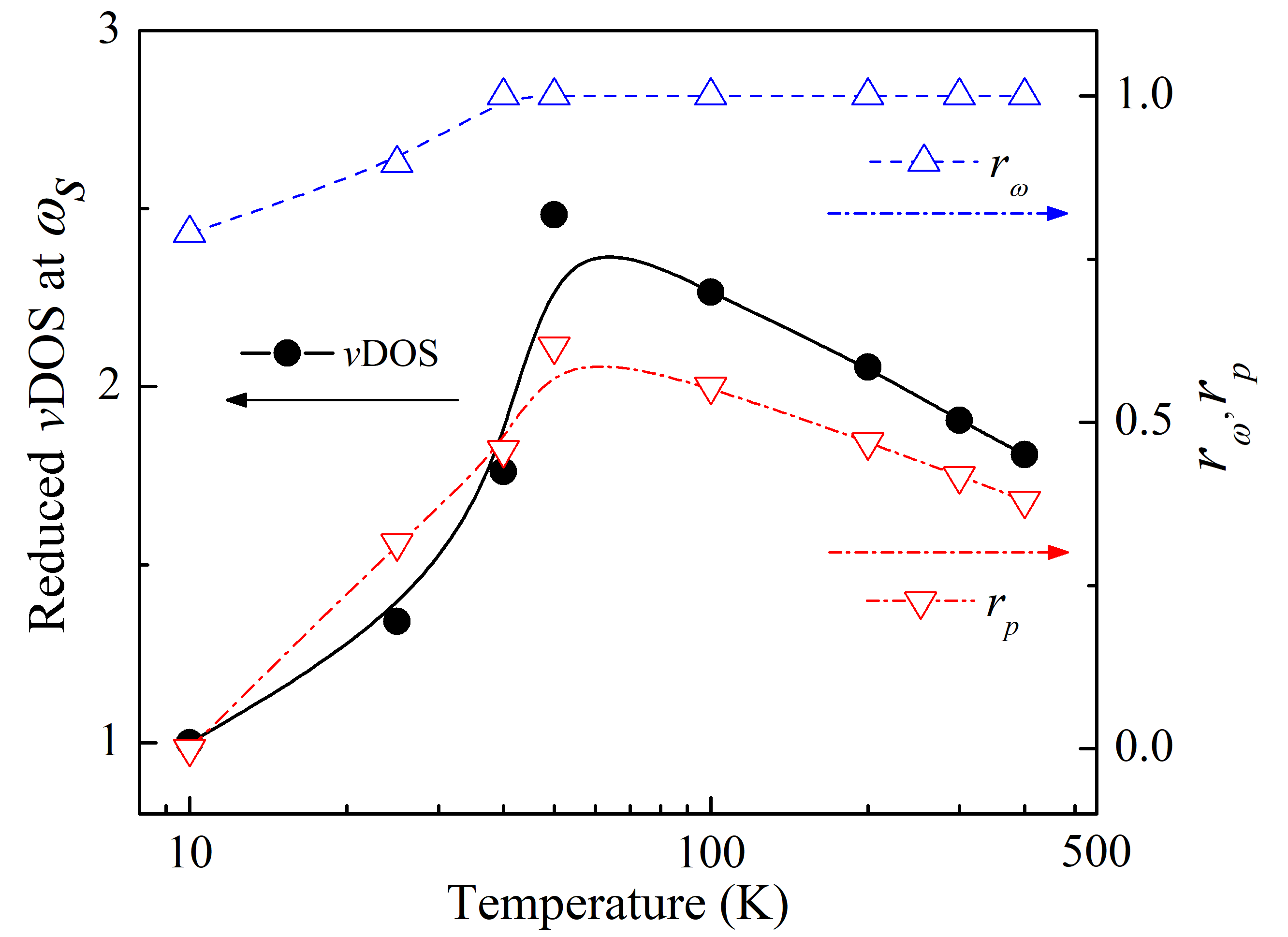}
\caption{Temperature effect on the synchronization of thermal phonons. The $k_{B}T$ reduced $vDOS$ at synchronization frequency $\omega _{s}$ (left-axis) and synchronization degree in frequency ($r _{\omega}$) and phase ($r _{p}$) (right-axis) as a function of temperature for the I1 system.
}
\label{fig6}
\end{figure}

From Eq. (\ref{eq_0}), we know that the nonequilibrium stationary state of self-synchronization demands a continuous external driving to offset the dissipative role of phonon-phonon scattering. In the thermal phonons system, the dissipated energy coming from the de-synchronization process is converted into the thermal energy of other modes, which can affect the vibrations and also the relative displacement of the pillars (see Fig. \,\ref{fig3}(d)). As observed in Fig. \,\ref{fig4}(b) and Fig. \,\ref{fig5}, the synchronization of thermal phonons that drives the potential energy minimization focuses the most of energies on the synchronization frequency $\omega _{s}$. Then the de-synchronization process is expected to further increase the potential energy between pillars. Finally, a energy balance is established between the phonon-phonon scattering resulted dissipative energy and the potential energy, respectively to the roles of $\Pi$ and $\Phi$ in Eq. (\ref{eq_0}). Thus, a nonequilibrium stationary state can be found after $\sim $ 500 ps, in which the self-synchronization of phase (Fig. \,\ref{fig3}) and mode energy ($\sim \tilde{T}$) at $\omega _{s}$ (Fig. \,\ref{fig5}) are converged.

The doping ratio also allows us to control the potential energy between pillars, by tuning the coupling constant $k$ in potential energy $\frac{1}{2}k\Delta x^{2}$, where $\Delta x$ is the relative displacement between pillars. Obviously, it can improve the balance between phonon-phonon scattering and potential energy. As shown in Fig. \,\ref{fig3}(a), the final self-synchronization degree in phase is significantly enhanced as doping ratio increase from 0.2 $\%$ to 0.4 $\%$.

\subsection{\label{sec:level1}Temperature effect on synchronization}

\begin{figure}[t]
\includegraphics[width=1.0\linewidth]{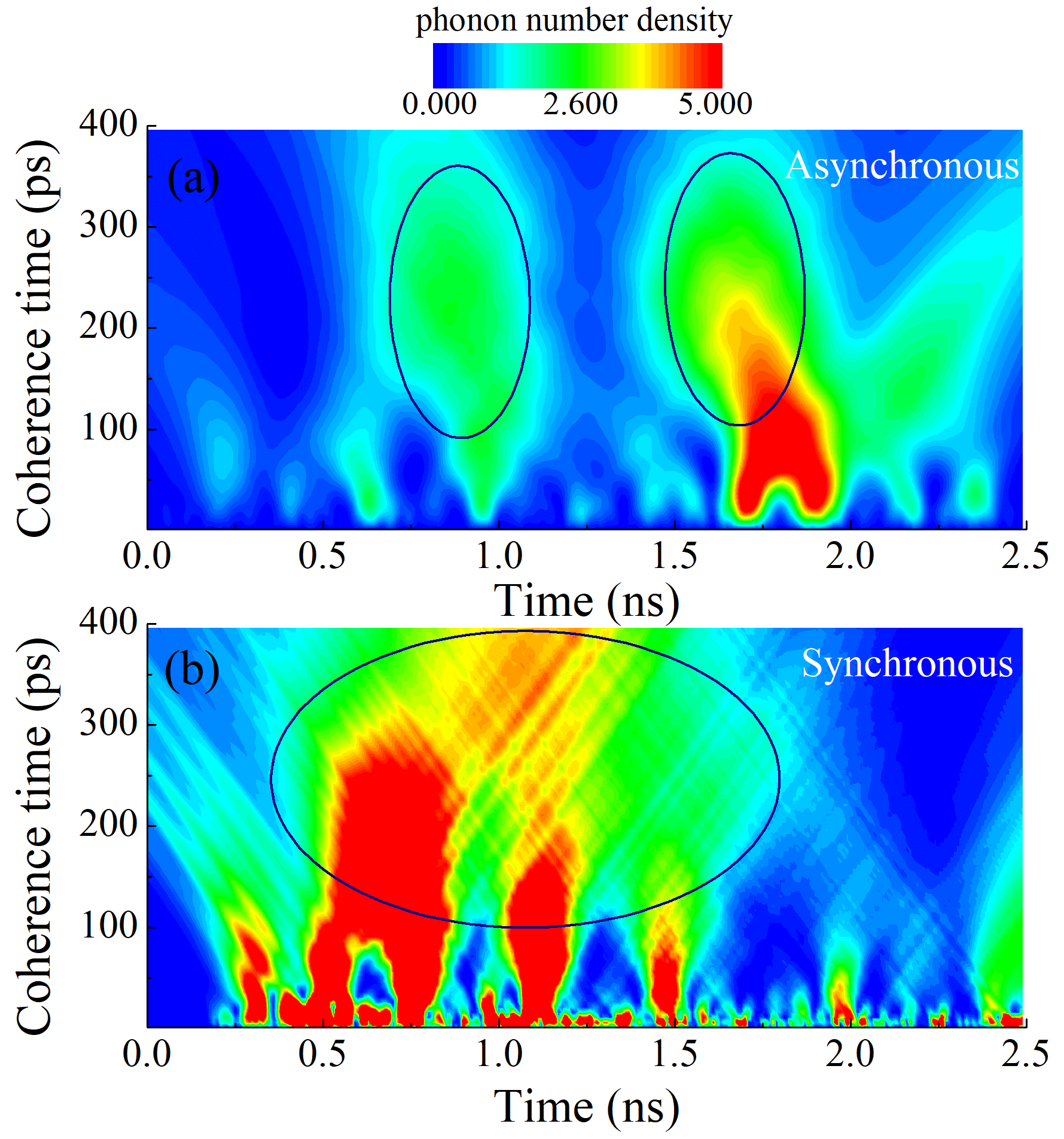}
\caption{Generation of coherent thermal phonons. Evolution time and coherence time dependent phonon number density for the thermal phonons at (a) asynchronous state and (b) synchronous state for the $\delta \Omega _{I}$ system at 100 K. The calculation for the synchronous state is in the I1 system and recorded after 600 ps. The circles of clouds indicate the emerging of thermal phonons.
}
\label{fig7}
\end{figure}

The temperature should have a significant effect on the self-synchronization of thermal phonons, by affecting the stationary state. As temperature increases, a competing relationship is raised, where temperature enhances the phonon-phonon scattering for de-synchronization but also elevates potential energy by increasing thermal fluctuation degree (displacement $\Delta x$). At low temperatures with weak phonon-phonon scattering, the increase of temperature should promote the synchronization by increasing the potential energy from thermal fluctuation, as manifested by the enhanced $r _{\omega}$ and $r _{p}$ in Fig. \,\ref{fig6}. However, as temperature continuously increases above 50 K, phonon-phonon scattering becomes significant and thus the phase synchronization degree $r _{p}$ is suppressed. In addition, the $k_{B}T$ reduced $vDOS$ at synchronization frequency $\omega _{s}$ is used to study the dynamics of synchronization energy, which exhibits the same same trend than phase synchronization $r _{p}$. The Fig. \,\ref{fig6} also evidences that a high synchronization degree in frequency is clearly much easier to achieve and more stable with temperature than a high $r _{p}$.

\section{\label{sec:level1}The generation of coherent thermal phonons}
 
Even in the partial synchronization state, the thermal phonons should be collectively locked in frequency and phase, in some degree, exhibiting a coherent state. Previous studies \cite{RN1636,RN1367,zhang2020generalized} demonstrated that the temporal coherence of thermal phonons can be analyzed by a wavelet transform approach. The calculation details can be found in Ref. \cite{SM}. Figure \,\ref{fig7} shows the calculated evolution time and coherence time dependent phonon number density $N\left ( t_{0},\tau _{s}^{c}  \right )$ of the I1 system. Compared to the phonon number density of a synchronous state, we find that thermal phonons in the asynchronous system are mostly distributed in the short coherence time regions. In addition, the clouds of phonon number density show rapid phonon creation and annihilation evolutions, exhibiting a short lifetime. After synchronizing, the thermal phonons exhibit a different coherence behavior in Fig.\,\ref{fig7}(b). The cloud of the $N\left ( t_{0},\tau _{s}^{c}  \right )$ distribution is extended both along evolution time and coherence time. Obviously, the emergence of the extended phonon cloud indicates the generation of new thermal phonons with long temporal coherence through the self-synchronization process. Moreover, the thermal phonons with a long coherence time also exhibit a long lifetime.

\section{\label{sec:level1}Conclusion}

Note that in bulk materials, thermal phonons can also be coupled through covalent bonding or long-range interactions \cite{RN1776,RN1227}. We hence expect self-synchronization of thermal phonons to also exist in other systems. In some extent, sufficient phonon-phonon scattering would suppress the synchronization degree and also its stability. On the other hand, coherent thermal phonons play an important role in thermal transport \cite{RN606,RN1230,RN975,RN184}. Therefore, self-synchronization could be a promising approach for tuning coherent thermal phonons and also thermal conductivity.

By performing MD simulations, we have demonstrated self-synchronization of thermal phonons due to the activation from thermal fluctuations in a doped silicon resonator system. We find that thermal phonons are spontaneously self-synchronized in frequency and phase. Phonon dynamics is analyzed based on the evolution process of frequency and phase synchronization. In addition, the results show, in comparison to synchronization in frequency, that the phase synchronization is harder to be achieve and to stabilize, which agrees well with the predictions of the Kuramoto model. Moreover, the synchronization degree of phase is sensitive to the intrinsic frequency difference and the coupling strength between oscillators. Small frequency differences and strong coupling would enhance the synchronization of thermal phonons. More interestingly, we find that there is a competing balance between energy dissipation resulting from phonon-phonon scattering and potential energy between resonators to maintain the stationary state of phase synchronization. Phonon-phonon scattering destroys the synchronized state but increases potential energy, while potential energy reversely feeds the synchronization of thermal phonons. This mechanism is further verified through the study of coupling strength and temperature effect. Eventually, we claim that the self-synchronization of thermal phonons is a new viewpoint for the generation of thermal phonons with long coherence time and lifetime. Our findings are likely to advance the understanding of thermal phonons dynamics from an unexpected perspective, and also promote the control of coherent thermal phonons.

\textit{Acknowledgments}---
This project is supported in part by the grants from the National Natural Science Foundation of China (Grant No. 11890703), Science and Technology Commission of Shanghai Municipality (Grant Nos. 19ZR1478600 and 18JC1410900 and 17ZR1448000), and the Fundamental Research Funds for the Central Universities (Grant No. 22120200069). This work is partially supported by CREST JST (No. JPMJCR19Q3). Z. Z. gratefully acknowledge financial support from China Scholarship Council.

\bibliographystyle{apsrev4-2}
\bibliography{library}

\end{document}



\title{Supplemental Material for ``Self-synchronization of thermal phonons at equilibrium"}

\author{Zhongwei Zhang}
\affiliation{Center for Phononics and Thermal Energy Science,\\
School of Physics Science and Engineering, Tongji University, 200092
Shanghai, PR China}
\affiliation{China-EU Joint Lab for Nanophononics, Tongji University, 200092 Shanghai, PR China}
\affiliation{Institute of Industrial Science, The University of Tokyo, Tokyo 153-8505, Japan}

\author{Yangyu Guo}
\affiliation{Institute of Industrial Science, The University of Tokyo, Tokyo 153-8505, Japan}

\author{Marc Bescond}
\affiliation{Laboratory for Integrated Micro and Mechatronic Systems, CNRS-IIS UMI 2820, University of Tokyo, Tokyo 153-8505, Japan}

\author{Jie Chen}
\email{jie@tongji.edu.cn}
\affiliation{Center for Phononics and Thermal Energy Science,\\
School of Physics Science and Engineering, Tongji University, 200092
Shanghai, PR China}
\affiliation{China-EU Joint Lab for Nanophononics, Tongji University, 200092 Shanghai, PR China}

\author{Masahiro Nomura}
\email{nomura@iis.u-tokyo.ac.jp}
\affiliation{Institute of Industrial Science, The University of Tokyo, Tokyo 153-8505, Japan}

\author{Sebastian Volz}
\email{volz@iis.u-tokyo.ac.jp}
\affiliation{China-EU Joint Lab for Nanophononics, Tongji University, 200092 Shanghai, PR China}
\affiliation{Laboratory for Integrated Micro and Mechatronic Systems, CNRS-IIS UMI 2820, University of Tokyo, Tokyo 153-8505, Japan}

\date{\today}

\maketitle

\section{\label{sec:level1}Phonon dispersion}

\begin{figure}[h]
\renewcommand\thefigure{S\arabic{figure}}
\includegraphics[width=0.40\linewidth]{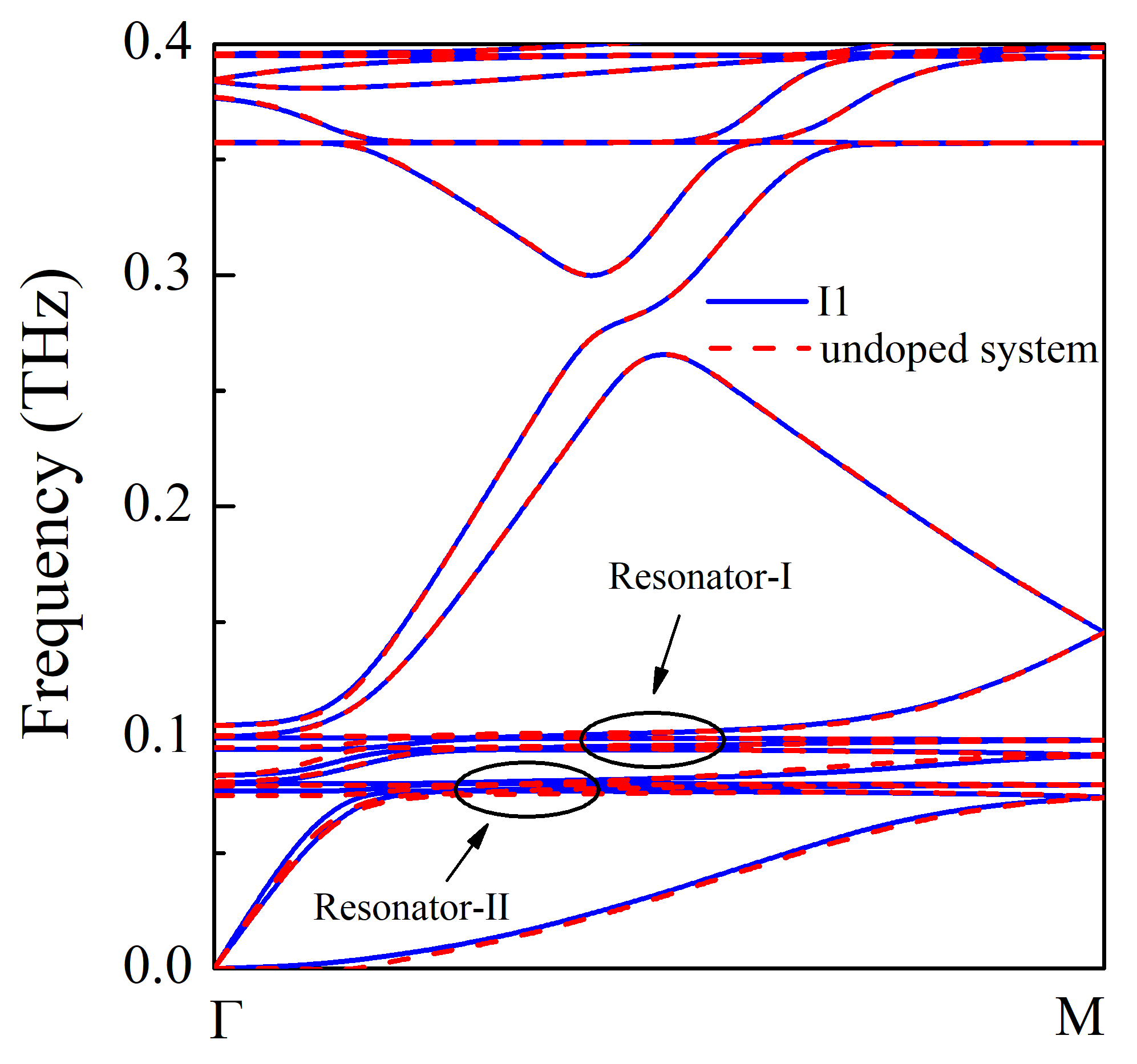}
\caption{Phonon dispersion from the harmonic approximation (0 K) of the I1 and undoped systems.
}
\label{bench}
\end{figure}

\section{\label{sec:level1}Wavelet transform method}

The temporal coherence of thermal phonons can be defined in the following basis

\begin{eqnarray}
\centering
\psi_{\omega_{\mathbf{k}s} ,t_{0},\Delta_{\mathbf{k}s} } \left ( t \right )=\pi ^{-\frac{1}{4}}\Delta_{\mathbf{k}s} ^{-\frac{1}{2}}e^{\left [ i\omega_{\mathbf{k}s} \left ( t-t_{0} \right ) \right ]}e^{\left [ -\frac{1}{2}\left ( \frac{t-t_{0}}{\Delta_{\mathbf{k}s}} \right) ^{2} \right ]},
\label{eq_7}
\end{eqnarray}

\noindent where $\omega_{\mathbf{k}s}$ is the angular frequency of mode ${\mathbf{k}s} $, and $\Delta_{\mathbf{k}s} $ defines the wavepacket duration. $t$ corresponds to the time variable, and $t_{0}$ to the position of highest amplitude in the wavepacket and also corresponds to the time evolution in the wavelet space. Inside the wavepacket, planewaves are in phase, the $\Delta_{\mathbf{k}s}$ term in Eq. (\ref{eq_7}) is thus a measure of the temporal coherence of thermal phonons. Here, we define the wavepacket full-width at half-maximum (FWHM) as the coherence time $\tau_{\mathbf{k}s}^{c}=2\sqrt{2ln2}\Delta_{\mathbf{k}s}$. The temporal coherence information of thermal phonons can be calculated from the wavelet transform as,

\begin{eqnarray}
\Lambda \left ( \omega_{\mathbf{k}s},t_{0},\tau_{\mathbf{k}s}^{c}\right )=\int \psi_{\omega_{\mathbf{k}s},t_{0},\tau_{\mathbf{k}s}^{c} } \left ( t \right )F\left (  t_{0}\right )dt,
\label{eq_8}
\end{eqnarray}

\noindent where $F\left (  t_{0}\right )$ denotes the time dependent dynamical quantity, which is chosen as the phonon modal velocity.

\begin{eqnarray}
\frac{1}{a}\sum_{b,l}^{a}\left [ \mathbf{\dot{u}}_{bl}\left ( t \right )\cdot \mathbf{e}^{\ast } _{b}\left ( \mathbf{k},s \right )\times exp\left ( i\mathbf{k}\cdot \mathbf{R}_{0l} \right )\right ],
\label{eq_9}
\end{eqnarray}

\noindent where $\mathbf{\dot{u}}_{bl}\left ( t \right )$ is the velocity of the $b$th atom in the $l$th unit cell at time $t$, $a$ is the number of cell, $\mathbf{e}^{\ast } \left ( \mathbf{k},s \right )$ complex conjugate of the eigenvector of mode ${\mathbf{k}s} $, and $\mathbf{R}_{0l} $ is the equilibrium position of the $l$th unit cell. $\omega_{\mathbf{k}s}$ refers to the resonance frequency in the undoped system and the synchronized frequency $\omega_{s}$ in the coupled system. At the frequency $\omega_{\mathbf{k}s}$, the time dependent phonon number of a given coherence time, here called phonon number density, $N\left ( t_{0},\tau _{s}^{c}  \right )$ can be calculated as $N\left ( t_{0},\tau _{s}^{c}  \right )=\frac{1}{2}m\left | \Lambda \left ( \omega_{s},t_{0},\tau _{s}^{c}  \right ) \right |^{2}/\hbar\omega_{s}$.